\documentclass{svjour2}

\smartqed

\usepackage{calc}
\usepackage{float}
\usepackage{graphicx}
\usepackage{bm}
\usepackage[percent]{overpic} 
\usepackage[colorlinks=true]{hyperref} 
\usepackage{amsmath}
\usepackage{amssymb}
\usepackage{latexsym}
\usepackage{multirow}
\usepackage{soul}

\usepackage{subcaption}
\usepackage[normalem]{ulem}

\newcommand{\be}{\begin{equation}}
\newcommand{\ee}{\end{equation}}
\newcommand{\bea}{\begin{eqnarray}}
\newcommand{\eea}{\end{eqnarray}}

\newcommand{\kommentar}[1]{}

\begin{document}
\title{Information Dimension of Stochastic Processes on Networks: 
Relating Entropy Production to Spectral Properties}
\date{\today}

\author{Oliver M\"ulken \and Sarah Heinzelmann \and Maxim Dolgushev}

\institute{
O. M\"ulken \at
              Physikalisches Institut, Universit\"at Freiburg, Hermann-Herder-Str.\ 3, D-79104 Freiburg, Germany \\
              \email{muelken@physik.uni-freiburg.de}\\
  \and
S. Heinzelmann \at
              Physikalisches Institut, Universit\"at Freiburg, Hermann-Herder-Str.\ 3, D-79104 Freiburg, Germany \\
 \and
M. Dolgushev \at
              Physikalisches Institut, Universit\"at Freiburg, Hermann-Herder-Str.\ 3, D-79104 Freiburg, Germany \\
              Institut Charles Sadron, Universit\'e de Strasbourg and CNRS, 23 rue du Loess, 67034 Strasbourg Cedex, France \\ 
              \email{dolgushev@physik.uni-freiburg.de}\\
              Tel.: +49-761-2037688\\
}

\date{Received: date / Accepted: date}


\maketitle

\begin{abstract} 
We consider discrete stochastic processes, modeled by classical master  equations, on networks. The temporal growth of the lack of information about the system is captured by its non-equilibrium entropy, defined via the transition probabilities between different nodes of the network. We derive a relation between the entropy and the spectrum of the master equation's transfer matrix. Our findings indicate that the temporal growth of the entropy is proportional to the logarithm of time if the spectral density shows scaling. In analogy to chaos theory, the proportionality factor is called (stochastic) information dimension and gives a global characterization of the dynamics on the network. These general results are corroborated by examples of regular and of fractal networks. 
\end{abstract}

\section{Introduction}

Many phenomena in physics, chemistry, biology or the social sciences can be modelled by
stochastic processes \cite{VanKampen1990}. The underlying dynamics can be described by Fokker-Planck, Langevin, or classical master equations. For the latter, one considers the transition probabilities between different possible (mesoscopic) states the system under investigation can be found in. This approach has been and still is very powerful in describing such diverse phenomena as first passage problems \cite{metzler2014first}, ergodicity breaking \cite{cherstvy2013anomalous}, or the population dynamics in reaction networks \cite{goutsias2013markovian}.

If the system of interest can be described by master equation with an underlying transfer matrix which is symmetric, one can diagonalize this matrix in order to obtain a spectral decomposition \cite{VanKampen1990}. Then, it has been found that many (long-time) quantities depend on the behavior of the spectral density for small eigenvalues. For many systems, it has been found that this part of the spectral density shows scaling, which then translates to scaling of quantities such as the return probability with an exponent called spectral dimension \cite{Alexander1982}.

In the long-time limit, many non-driven systems will reach a stationary state, which often is the thermodynamic equilibrium state \cite{VanKampen1990}. The way in which this state is approached is determined by the temporal behavior of the transition probabilities. Since the equilibrium state is also the state with the largest entropy, one can also use the entropic measures to describe the temporal evolution of the system. This is also an important aspect in stochastic thermodynamics, see for instance \cite{seifert2012stochastic,esposito2012stochastic,van2015ensemble}

In this paper, we discuss the temporal behavior of different time-dependent entropies, defined on the basis of the transition probabilities. As we are going to show, the temporal increase of these entropies can be related to the spectral dimension, if the underlying transfer matrix allows to define it. This enables us to give an analytical expression for the so-called (stochastic) information dimension, introduced on the basis of numerical calculations for percolation lattices \cite{argyrakis1987information,pitsianis1989information}.

\section{Entropy of Stochastic Processes}

We consider systems constituted by a set of possible states $|j\rangle$ ($j=1,\dots,N$) and with a stochastic dynamics described on the basis of a master equation of the following type \cite{VanKampen1990}
\be
\dot p_{kj}(t) = \sum_{l=1}^N T_{kl} \, p_{lj}(t).
\label{eq.master}
\ee
Here, $p_{kj}(t)$ denote the transition probabilities to go from state $|j\rangle $ to state $|k\rangle$ and the matrix ${\bm T}$ is the so-called transfer matrix with elements $T_{kj}$. We typically assume localized initial conditions of the type $p_{kj}(0)=\delta_{kj}$, where $\delta_{kj}$ is the Kronecker delta.
The matrix ${\bm T}$ is real. If it is also symmetric, one can use the 
spectral decomposition ${\bm T} | \phi_n \rangle = - \lambda_n | \phi_n \rangle$, with the eigenstates $| \phi_n \rangle$ and where the eigenvalues $\lambda_n\geq0$, to express the
formal solution of Eq.~(\ref{eq.master}) by
\bea
p_{kj}(t) &=& \langle k | \exp{{\bm T}t} | j \rangle = \sum_{n=1}^N e^{-\lambda_n t} \langle k | \phi_n \rangle \langle \phi_n | j \rangle 
\nonumber \\
&=& \sum_\lambda e^{-\lambda t} \underbrace{\sum_{m=1}^{D(\lambda)} \langle k | \phi_m \rangle \langle \phi_m | j \rangle}_{\equiv D_{kj}(\lambda)}, 
\label{eq.pkj}
\eea
where $D(\lambda)$ is the degeneracy of eigenvalue $\lambda$ and $D_{kj}(\lambda)$ is a state-dependent weight depending on $D(\lambda)$.

The knowledge of $p_{kj}(t)$ is in principle sufficient to determine all dynamic and static properties of the considered system. For instance, the average probability to return to (or remain at) the initial state, $\overline p(t)$, allows for a global statement about, say, transport efficiency \cite{Muelken2006,Muelken2011}. Using Eq.~(\ref{eq.pkj}), one has
\be
\overline p (t) \equiv \frac{1}{N} \sum_j p_{jj}(t) = 
\frac{1}{N} \sum_\lambda e^{-\lambda t} D(\lambda),
\label{eq.pavg}
\ee
which, in the limit $N\to\infty$ with a continuous spectrum, becomes
\be
\overline p (t) = \int d\lambda \, \rho(\lambda) \, e^{-\lambda t},
\label{eq.pavg-cont}
\ee
where $\rho(\lambda)$ is the spectral density \cite{Alexander1982}.

\subsection{Stochastic Entropy}

Another quantity to characterize the stochastic process is the non-equilibrium (Shannon) entropy \cite{VanKampen1990}, given by
\be
S_j(t) = - \sum_{k=1}^N p_{kj}(t) \ln [p_{kj}(t)],
\label{eq.entropy}
\ee
which can still depend on the initial state $|j\rangle$. Averaging over all possible initial states yields
\be
\overline S(t) = \frac{1}{N} \sum_{j=1}^N S_j(t).
\label{eq.entavg}
\ee
For localized initial states, $p_{kj}(0) = \delta_{kj}$, one has $S_j(0)=\overline S(0) =0$. In the limit $t\to\infty$, the system approaches its equilibrium distribution $\lim_{t\to\infty} p_{kj}(t) \equiv p_{\rm eq}$, such that 
\be
S_{\rm eq} \equiv \lim_{t\to\infty} S_j(t)= \lim_{t\to\infty} \overline S(t) = - N p_{\rm eq} \ln p_{\rm eq}.
\ee
For undirected finite networks with $N$ nodes, one has $p_{\rm eq}=1/N$, thus, $S_{\rm eq}=\ln N$.

\subsection{Mean Field Approach}

For intermediate times, the importance of the initial condition has decreased, such that we assume averaged transition probabilities
\be
p_{kj}(t) \approx 
\begin{cases}
\overline p (t) & \text{for} \,\, k=j \\
{\displaystyle\frac{1-\overline p(t)}{N-1}} & \text{else.}
\end{cases}\label{eq.pmf}
\ee
As can be inferred from Eq.~(\ref{eq.pmf}), a rather homogeneous character of the nodes is assumed. We then obtain the {\it mean-field} averaged entropy
\bea
\overline S_{\text{mf}} (t) 
&\equiv&
- \overline p(t) \ln[\overline p(t)] 
- [1-\overline p(t)] \ln\left[ \frac{1-\overline p(t)}{N-1} \right] \nonumber \\
& = &
- \ln [\overline p(t)] + [1-\overline p(t)] 
\ln\left[\overline p(t) \frac{N-1}{1-\overline p(t)} \right] ,
\label{eq.smf}
\eea

The most significant contribution for large intermediate/transient times comes from the first term, resulting in
\be
\overline S_{\text{mf}} (t) \sim - \ln[\overline p(t) ] \geq 0.
\label{eq.mfapprox}
\ee
This equation is especially intriguing, since it relates the dissipative entropy growth to the spectrum of the underlying transfer matrix via the spectral density $\rho(\lambda)$, which determines the behaviour of $\overline p(t)$.
Furthermore, $-\ln[\overline p(t)]$ recovers the limit of $\overline S(t)$ for $t\to\infty$, namely, $-\lim_{t\to\infty} \ln[\overline p(t)] =\ln N$. We note that one also has $-\ln[\overline p(0)] = 0$, such that we find $-\ln[\overline p(t)] \in [0,\ln N]$.

\subsection{Stochastic Information Dimension}

Entropy can be viewed as a measure of the lack of information about the system. As such, it can be used to define, for an infinite system, the stochastic analog of the so-called information dimension known in chaos theory \cite{OttChaos}:
\be
d_i \equiv \lim_{t\to\infty} \frac{\overline S(t)}{\ln t}.
\ee 
Consequently, we define in the mean field approximation, on the basis of Eq.~(\ref{eq.mfapprox}), and for an infinite system
\be
d_{i,{\rm mf}} \equiv - \lim_{t\to\infty}  \frac{\ln[\overline p(t)]}{\ln t}.
\ee
For finite networks we consider the behavior for long intermediate/transient times before saturation to the equilibrium value $S_{\rm eq}$ sets in.

Thus, the behavior of $\overline p(t)$ for large $t$ gives the most significant contributions. By means of Eqs.~(\ref{eq.pavg}) and~(\ref{eq.pavg-cont}), one finds that in this case the behavior of $\rho(\lambda)$ for small values of $\lambda$ is important, since for long times all contributions for large values of $\lambda$ have already died out. A large variety of networks show scaling behavior of $\rho(\lambda) \sim \lambda^\nu$, where $\nu=d_s/2 -1$ is related to the so-called spectral dimension $d_s$ \cite{Alexander1982}.\footnote{Note that other definitions of the spectral dimension(s) based on $p_{jj}(t)$ and $\overline p(t)$ exist \cite{agliari14,agliari16,burioni05}.} It is straightforward to show that then \cite{Alexander1982,Muelken2006,Muelken2011}
\be
\overline p(t) \sim t^{-d_s/2} \qquad \text{for} \, t\gg1,
\ee
and
\be
d_{i,{\rm mf}} = d_s/2.
\label{eq.dimfds}
\ee
In this way $2d_{i,{\rm mf}}$ resembles the definition of the so-called average spectral dimension \cite{burioni05}. However, there are  networks with an inhomogeneous distribution of states \cite{agliari14,burioni05}. One of such structures is considered in the Appendix.

It is not possible to derive in general such a relation for $d_i$ itself, since for this it would be necessary to have the knowledge of the whole probability distribution $p_{kj}(t)$. However, we will show in the following examples that the same relation also holds for $d_i$.

\section{Examples}

\subsection{Regular Networks}

We start by considering regular networks. In its most basic form, this is a line or ring of $N$ nodes which have $d_s=1$. For the ring we can employ Bloch's theorem for the eigenstates \cite{Ashcroft}, such that 
\be
p_{kj}(t) = \frac{1}{N} \sum_{n=1}^N e^{-\lambda_n t} e^{i\theta_n (k-j)},
\ee
where $\theta_n = 2 \pi n/N$ and $i$ it the imaginary unit. The eigenvalue $\lambda_n = 2 - 2 \cos \theta_n$. In the limit $N\to\infty$, one has
\be
p_{kj}(t) = e^{-2t} \int\limits_0^{2\pi} d\theta \, e^{2 t \cos\theta} e^{i\theta(k-j)} = 2\pi e^{-2t} I_{(k-j)}(2t),
\ee
where $I_{(k-j)}(2t)$ is the modified Bessel function of the first kind \cite{abramowitz}.
For long time one has
\be
I_{(k-j)}(2t) = (4\pi t)^{-1/2} e^{2t},
\ee
which yields
\be
p_{kj}(t) = (t/\pi)^{-1/2} \qquad \text{for all} \, \,  k, j.
\ee
Thus, here we find that the entropy $S_j(t)$ for long times becomes independent of $j$ and reads
\be
S_j(t) = \overline S(t) = \frac{1}{2} \ln [t/\pi].
\ee
Thus we have the logarthmic increase of $S_j(t)$ as well as for $\overline S(t)$ with 
\be
d_i = d_s/2 = 1/2.
\ee

Obviously, the same information dimension, $d_{i, {\rm mf}}=d_i$, is obtained from the mean-field approach, since the transition probabilities $p_{kj}(t)$ become state independent for long times. In particular, we obtain
\be
\overline p(t) = e^{-2t} \int\limits_0^{2\pi} d\theta \, e^{2 t \cos\theta} = 2\pi e^{-2t} I_{(0)}(2t)
\ee
and therefore, for long times, 
\be
\overline S_{\rm mf} = \frac{1}{2} \ln [t/\pi].
\ee

We note that this analysis is easily extended to regular networks of higher dimension $d$ where one finds $d_s=d$. In this case the $d$-dimensional regular network is a direct product of $d$ one-dimensional regular networks.

\subsection{Fractal Networks}

\subsubsection{Generic solution of Master Equation}

For some fractal networks, it has been shown by comparison to numerical computations, that the solution of the master equation is of
stretched exponential form \cite{Blumen1991}. Exemplarily, so-called Sierpinski gaskets have a solution of the form
\be
p_{kj}(t) \sim t^{-d_s/2} \exp\left( -a \xi_{k-j}^\nu \right),
\ee
where $\xi_{k-j} = |k-j| t^{-d_s/2d_f}$, $d_f$ is the fractal dimension of the network and $\nu$ is related to the walk dimension $d_w = 2d_f/d_s$. 
Inserting this into Eqs.~(\ref{eq.entropy}) and (\ref{eq.entavg}) yields
\be
S_j(t) 
\sim \frac{d_s}{2} \ln t \sim \overline S(t),
\ee
and also confirms Eq.~(\ref{eq.mfapprox}), i.e., $\overline S_{\text{mf}}(t) \sim (d_s/2) \ln t$.

\subsubsection{Vicsek fractals}

\begin{figure}[h]
\centerline{\includegraphics[width=0.7\columnwidth]{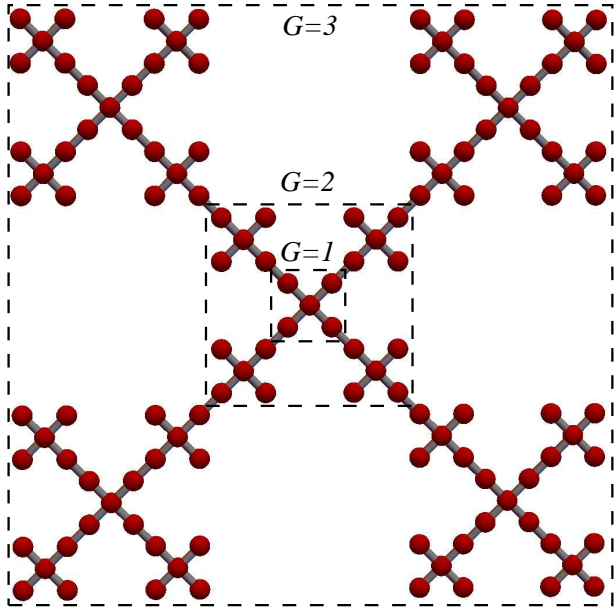}}
	\caption{(color online)  Sketch of a Vicsek fractal of generation $G=3$ with $f=4$. 
}
\label{fig.vicsek}
\end{figure}

As one particular example, we chose the Vicsek fractals \cite{blumen2003dynamics}. These are tree-like structures which are built iteratively, starting from a simple star with $f$ arms (generation $G=1$); to each arm one attaches $f$ equivalent stars, resulting in a larger structure (generation $G=2$); $f$ of such structures are then attached to the central structure (generation $G=3$); etc., see Fig.~\ref{fig.vicsek} for a Vicsek fractal of generation $G=3$ with $f=4$. The total number of nodes in generation $G$ is $N_G=(f+1)^G$.

While it is straightforward to calculate numerically the different entropies, in particular $\overline S(t)$ and $\overline S_{\rm mf}(t)$, we can also compare this to analytic expressions based on the knowledge of the spectrum. It is possible to obtain the eigenvalues of the transfer matrix iteratively, see \cite{blumen2003dynamics} for details. Thus, for finite $G$, we can compare the full numerical computation of $\overline S(t)$ to the leading term of $\overline S_{\rm mf}(t)$ given by Eq.~(\ref{eq.mfapprox}), which only depends on the spectrum.
Since the iterative computation of the spectrum also allows to obtain the spectral dimension \cite{blumen2003dynamics}
\be
d_s = \frac{2\ln(f+1)}{\ln(3f+3)}
\ee
we can check the validity of Eq.~(\ref{eq.dimfds}).

\begin{figure}[h]
\centerline{\includegraphics[width=0.9\columnwidth]{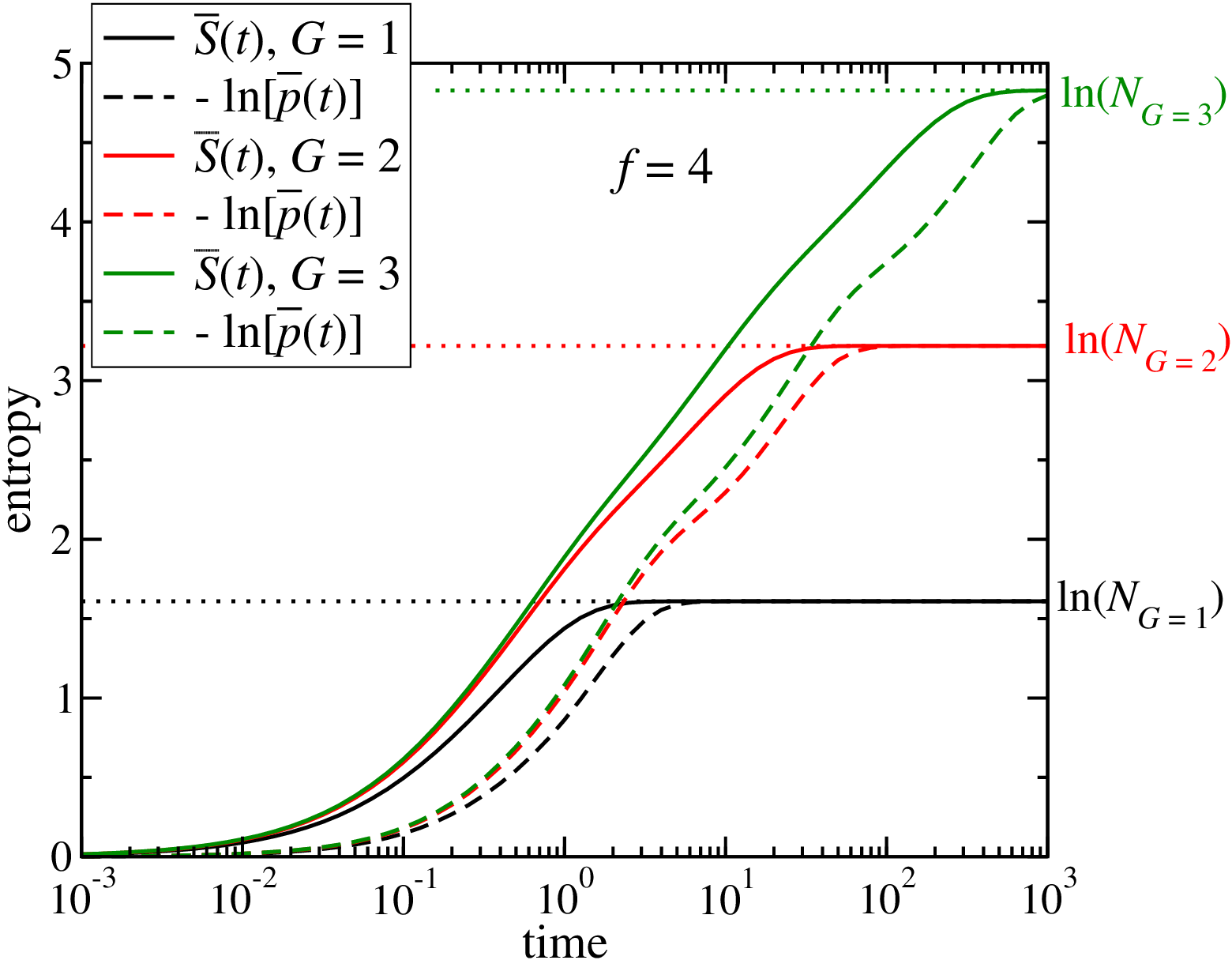}}
	\caption{(color online) Entropy $\overline S(t)$ and the leading term of $\overline S_{\rm mf}(t)$, Eq.~(\ref{eq.mfapprox}), for $f=4$ and $G=\{1,2,3\}$. 
}
\label{fig.vicsek1}
\end{figure}

Figure \ref{fig.vicsek1} shows the comparison of $\overline S(t)$ with the leading term of $\overline S_{\rm mf}(t)$, Eq.~(\ref{eq.mfapprox}), for $f=4$ and $G=\{1,2,3\}$. One clearly notices the step-like structure of both quantities with increasing $G$. Moreover, one also finds a similar behavior with increasing $t$, albeit a shift to longer times for the leading term of $\overline S_{\rm mf}(t)$. This already indicates that the qualitative temporal increase of the entropy is mainly determined by the spectrum. In addition, also the step-like structure can be attributed to spectral properties, since the spectrum of Vicsek fractals is not smooth but rather separated into distinct segments corresponding to eigenvalues appearing with increasing $G$, see also Ref.~\cite{blumen2003dynamics} for a more detailed analysis of the spectrum. Step-like structures have also been observed for different quantities depending only on the spectrum, such as the storage and loss moduli for fractal polymers \cite{mielke2016relaxation}.

\begin{figure}[h]
\centerline{\includegraphics[width=0.9\columnwidth]{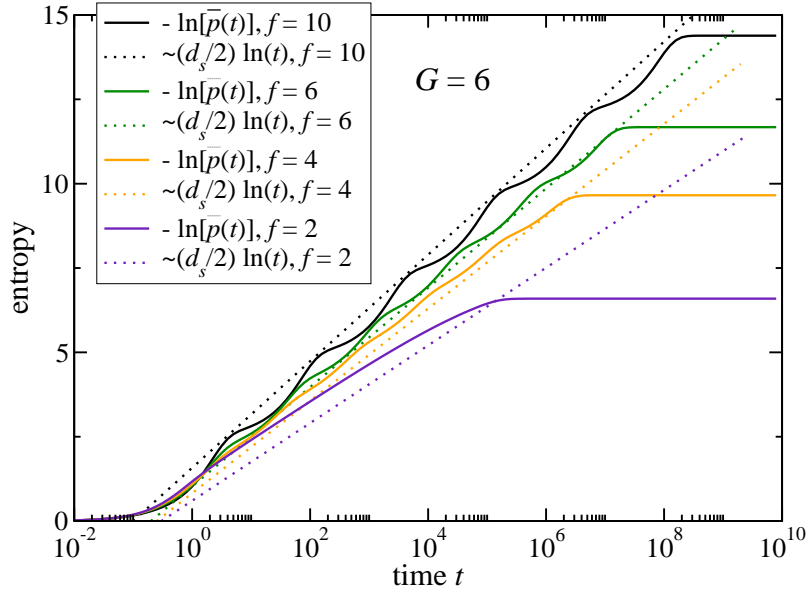}}
	\caption{(color online) Different approximations of entropy, $-\ln[\overline p(t)]$ and $(d_s/2) \ln(t)$, for $G=6$ and $f=\{2,4,6,10\}$. 
}
\label{fig.vicsek2}
\end{figure}

In Fig.~\ref{fig.vicsek2}, we compare $-\ln[\overline p(t)]$ with $(d_s/2) \ln(t)$ for $G=6$ and different $f=\{2,4,6,10\}$. Again there is a shift to longer times for the  $(d_s/2) \ln(t)$. However, the step-like increase of  $-\ln[\overline p(t)]$ is well approximated by the linear increase with $\ln(t)$. Thus, also for Vicsek fractal we conclude that, indeed a definition of an information dimension is reasonable, with $d_i = d_s/2$.

\subsection{Counter example}\label{counter_example}

A logarthmic-in-time increase of entropy will not always appear. As we have demonstrated above, the increase in entropy is related to the temporal behavior of $\overline p(t)$, which, in turn, is related to the spectral density $\rho(\lambda)$. Only if $\rho(\lambda)$ has a continuous power-law part for small values of $\lambda$, one finds the logarithmic increase. There are examples for which such a continuous part does not exist (at least for small values of $\lambda$). For instance, take the complete graph, where every node is directly connected by single bonds to the remaining $(N-1)$ nodes. The spectrum consists only of two values $\lambda_1=0$ with $D(\lambda_1)=1$ and $\lambda_2=N$ with $D(\lambda_2) = N-1$. Thus, the only time-scale involved is $Nt$. Using in Eq.~(\ref{eq.pkj}) that 
\be
\sum_{m=2}^{D(\lambda_2)} | \phi_m \rangle \langle \phi_m | = \underbrace{\sum_{m=1}^N | \phi_m \rangle \langle \phi_m |}_{=\bm 1} - | \phi_1 \rangle \langle \phi_1 |,
\ee 
with 
\be
|\phi_1 \rangle = \frac{1}{\sqrt{N}} \sum_{l=1}^N | l \rangle,
\ee
we find that
\be
p_{kj}(t) = \frac{1}{N} \Big[ 1- \underbrace{e^{-Nt} \Big(1 - N \delta_{kj} \Big)}_{\ll 1 \quad \text{for} \, \, t\gg1} \Big]
\ee
yielding
\bea
S_j(t) 
&=& - \frac{1 + (N-1) e^{-Nt}}{N}\ln\left[\frac{1 + (N-1) e^{-Nt}}{N}\right] \nonumber \\
&& - (N-1)\frac{1-e^{-Nt}}{N}\ln\left[\frac{1-e^{-Nt}}{N}\right]
\eea
Since all nodes are equivalent, one has $\overline S(t) = S_j(t)$.
For long times and large $N$, this results in 
\be
\overline S(t) \approx \ln N - N e^{-2Nt},
\ee
thus an exponential convergence to the equilibrium value. 

For the mean-field expression, we obtain
$\overline S_{\rm mf}(t) \approx \overline S(t)$. However, for the leading term of $\overline S_{\rm mf}(t)$ given in Eq.~(\ref{eq.mfapprox}) we get
\bea
- \ln[\overline p(t)] 
&=& - \ln[1/N + e^{-Nt}] \nonumber \\
&\approx& \ln N - N e^{-Nt} \, \, \text{for} \, \, t\gg1
\eea
which also converges to the equilibrium value of $\ln N$ but in a different functional form than $\overline S(t)$ and $\overline S_{\rm mf}(t)$.

\section{Conclusion and Outlook}

For stochastic processes modelled by a classical master equation, we have related the logarithmic growth of the time-dependent entropy to the spectral properties of the underlying transfer matrix. Our findings indicate that a direct relation only exists if the spectral density shows scaling for small eigenvalues, namely, when one can define a spectral dimension. Then it is possible to define a stochastic information dimension which happens to be half the spectral dimension. We have corroborated our general statements by examples of regular and of fractal networks. In addition we have shown the different temporal growths for systems whose spectrum does not allow to define a spectral dimension.

Finally, we would like to mention a relation to statistical physics, which we think is important, but the analysis of which goes beyond the scope of this paper:
Since entropy $S_j(t)$ will increase with time, so will $\overline S(t)$ as well as $\overline S_{\rm mf}(t)$. Therefore, we use
their time-derivates to define the entropy production rate(s):
\bea
\dot S_j(t) &=& - \sum_{k=1}^N \dot p_{kj}(t) \ln [p_{kj}(t)], \\
\dot{\overline S}(t) &=& \frac{1}{N}\sum_j \dot S_j(t), 
\eea
and, with Eq.~(\ref{eq.mfapprox}), one further obtains
\be
\dot{\overline S}_{\text{mf}}(t) \sim - \frac{\dot{\overline p}(t)}{\overline p(t)} = \sum_\lambda \lambda \, \underbrace{\frac{\rho(\lambda) e^{-\lambda t}}{\sum_\lambda \rho(\lambda) e^{-\lambda t}}}_{\equiv p_t(\lambda)} \equiv \langle \lambda \rangle_t.
\ee
By realizing that  $\langle \lambda \rangle_t$ is reminiscent of an average value with a time-dependent probability distribution $p_t(\lambda)$, one obtains a relation between the entropy production and the spectrum of the transfer matrix ${\bm T}$.
We believe that this is certainly an issue worth of further investigation, in particular due to its striking resemblance of expressions known from equilibrium statistical physics, such as the average energy (identified with $\lambda$) at a given inverse temperature (identified with $t$) in the canonical ensemble \cite{reichl1980modern}. 

We close by mentioning that an extension to open quantum systems might also be feasible. Based on, say, quantum master equations for the reduced density operator, one can define an extension of the stochastic information dimension via the temporal increase of the von-Neumann entropy \cite{schijven2014information}.

\begin{acknowledgements}

We thank Alex Blumen for fruitful discussion and valuable comments. M.D. acknowledges the support through Grant No. GRK 1642/1 of the Deutsche Forschungsgemeinschaft. 

\end{acknowledgements}


\section*{Appendix: $NT_D$ graph}

In this Appendix we extend our results by considering a structure with a non-monotonous density of states $\rho(\lambda)$. 

\begin{figure}[!ht]
\centerline{\includegraphics[width=0.99\columnwidth]{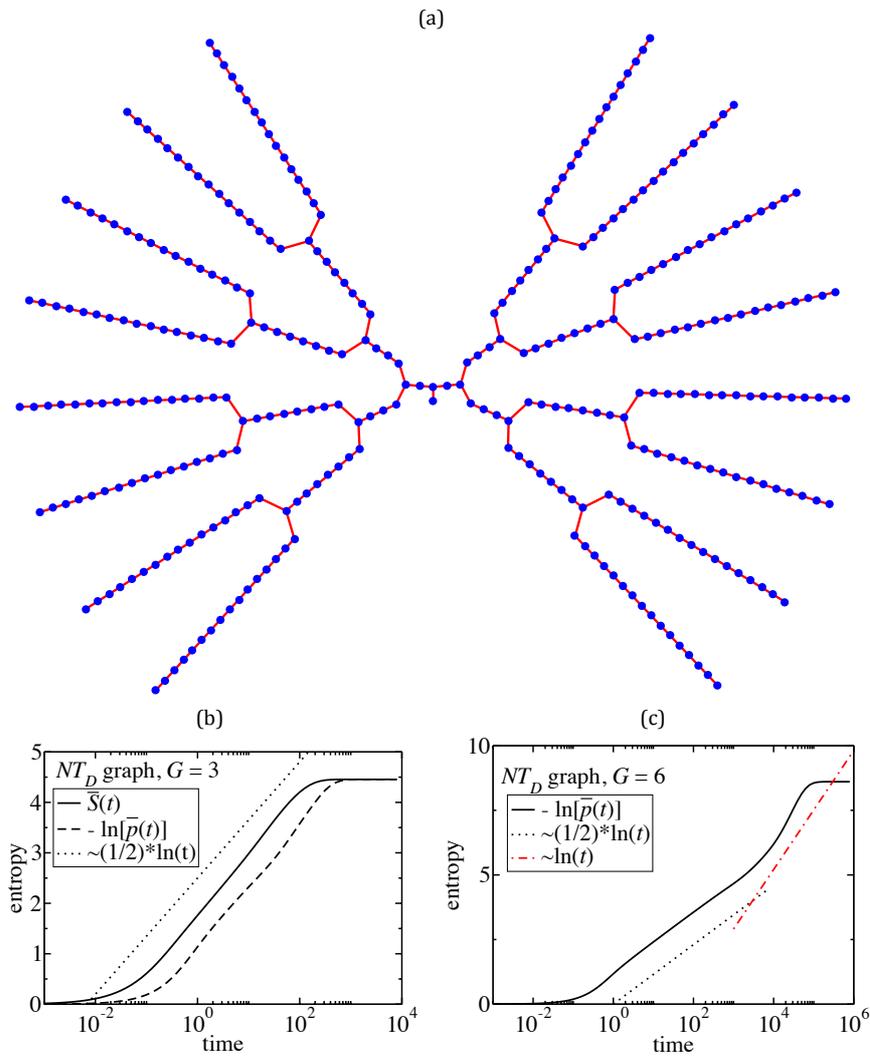}}
	\caption{(color online)  (a) Sketch of an $NT_D$ graph of generation $G=4$ and $k=2$. (b) Entropy $\overline S(t)$ and the leading term of $\overline S_{\rm mf}(t)$, Eq.~(\ref{eq.mfapprox}), for $NT_D$ graph of $k=2$ and $G=3$. (c) The behavior of $-\ln[\overline p(t)]$, $(d_s^{\mathrm{lin}}/2) \ln(t)$ and $(d_s^{\mathrm{br}}/2) \ln(t)$, for $G=6$ and $k=2$.
}
\label{fig.ntd}
\end{figure}

In Fig.~\ref{fig.ntd}(a) we illustrate the construction of the so-called $NT_D$ graph \cite{burioni94,burioni95,burioni05}. At every iteration $G$ to each end-node of these trees $k$ linear chains of length $2^G$ are attached. In this way, the Laplacian spectrum of the $NT_D$ graphs is dominated by the behavior of the linear chains, whose spectral dimension is $d_s^{\mathrm{lin}}=1$. On the other hand, for these trees the states corresponding to the smallest eigenvalues are described by the relaxation of the branches as whole (similarly as for dendrimers \cite{cai97,gotlib02,gurtovenko03,fuerstenberg12}). As has been found in Refs.~\cite{burioni94,burioni95}, the related spectral dimension is given by $d_s^{\mathrm{br}}=1+\log k/\log 2$. As we proceed to show, both aspects of the spectrum $\{\lambda\}$ are reflected in the temporal growth of the entropy.

First, in Fig.~\ref{fig.ntd}(b) we show that, as for Vicsek fractals, the leading term $-\ln[\bar{p}(t)]$ determines the temporal behavior of the entropy $\bar{S}(t)$ for the $NT_D$ graphs. We observe a scaling of the linear chains, i.e. $\bar{S}(t)\sim(1/2)\ln(t)$. Increasing generation $G$ leads to an appearance of lower and lower $\lambda$'s that get separated from the (continuous) spectrum of the linear chain. This leads to a change in the behavior of the term $-\ln[\bar{p}(t)]$ for longer times, see Fig.~\ref{fig.ntd}(c). Thus, different parts of the spectrum $\{\lambda\}$ translate their behavior to the time-dependent entropy $\bar{S}(t)$.


\end{document}